\documentclass[10pt,preprint,showpacs,twocolumn]{revtex4}
\usepackage{amsfonts}
\usepackage{graphicx}
\usepackage{amsmath}
\usepackage{amsthm}
\usepackage[colorlinks=true, citecolor=blue, urlcolor=blue ]{hyperref}
\usepackage{graphicx}
\usepackage{textcomp}
\long\def\/*#1*/{}

\begin{document}
\title{ Geometry of quantum state space and entanglement}

\author{Pratapaditya Bej}
\email{pratap6906@gmail.com}
\author{Prasenjit Deb}
\email{devprasen@gmail.com}
\affiliation{Department of Physics and Center for Astroparticle Physics and Space Science, Bose Institute, Bidhan Nagar
Kolkata - 700091, India.}

\begin{abstract}
Recently, an explicit relation between a measure of entanglement and a geometric entity has been reported in
\textcolor{blue}{Quantum Inf. Process. (2016) 15:1629-1638}. It has been shown that if a qubit gets entangled with another ancillary qubit then negativity, upto a constant factor, is equal to square root of a specific Riemannian metric defined on the metric space corresponding to the state space of the qubit. In this article, we consider the different class of bi-partite entangled states and show explicit relation between two measures of entanglement and Riemannian metric.

\end{abstract}
\maketitle

\section{Introduction}
Nowadays, geometric tools are often used in quantum information theory because of the fact that these tools provide advantage to find out less trivial and robust physical constraints on physical systems. Among such various tools, \emph{differential geometry} is an important one. Before quantum information it was applied in classical information theory. As a result of which a new discipline, called \emph{Information Geometry} 
emerged  and it got maturity through the works of Amari,Nagaoka and other mathematicians in the 1980s\cite{Amari}. Initially,the goal of Information geometry was to understand the interplay between the information-theoretic quantities and the geometry of probability space by constructing a Riemannian space corresponding to probability space. Later, Morozova and \u{C}encov \cite{Cencov} extended the geometric formulation of probability space to quantum setting by proposing Riemannian metrics on the space of density matrices. Their study gradually progressed through the works 
of Petz and other authors \cite{Pet1,Pet2,Pet3,Pet4,Isola1,Isola2}. The monotone Riemannian metric corresponding to Wigner-Yanase-Dyson skew-information \cite{Wigner} was found out in \cite{Pet4} which expresses the relation between geometry of space and an information-theoretic quantity of great importance. Geometric distances(metrics) are also shown to be useful in quantum  state discrimination problem \cite{Wootters,Caves}. In\cite{Pinto} the authors have demonstrated that a lower bound for quantum coherence measure can be found out using Riemannian monotone metric. 

\paragraph*{}
On the other hand, quantum correlation is a resource in quantum information processing. Though there are different aspects of quantum correlations, entanglement and discord are the two aspects which have been extensively studied. However, till date, quantum correlation is not fully understood. So, the study of quantum correlations demands importance in quantum information theory. Here we consider entanglement because of the fact that all the measures of entanglement are monotonic in nature. Quantum entanglement\cite{Schrodinger,Werner,Horodecki} is one of the bizzare phenomena exhibited by composite quantum systems. It is a resource for quantum information processing tasks, such as teleportation \cite{Tele}, dense coding \cite{Dense}, quantum cryptography \cite{Crypto}, state merging\cite{SM}, quantum computation and many more. A composite quantum system $\rho_{AB}$ is said to be entangled if it can not be written as $\rho_{AB}= \sum_i~p_i \rho_A^i\otimes\rho_B^i$, where $p_i$ are probabilities, $\rho_A$ and $\rho_B$ are respectively the desity matrices of subsystem $A$ and $B$. If the subsystems are two-level quantum states then these are termed as \emph{qubits} \cite{Nielsen} in analogy with classical bits. Qubits are the fundamental units in quantum information theory. 
\paragraph*{}
Entanglement is the most studied form of quantum correlations. However, geometry of quantum state space has not been applied in the study of quantum entanglement till date. Therefore, study of quantum correlations using geometry of quantum state space will be an interesting area of research. To begin with, one may address the problem of finding unique Riemannian metrics corresponding to different measures of entanglement. Interestingly, a problem close to this has recently been addressed in\cite{pd}. An explicit relation between negativity, a measure of entanglement, and monotone Riemannian metric was established in that article. The author considered entanglement generation between two qubits and calculated the negativity of the generated entangled state to establish the relation bewteen a geometric entity(Riemannian metric) and an entanglement measure(negativity). A specific unitary operation was considered which can create entanglement between two qubits, initially in a product form. However, one question remained unanswered in that paper: \emph{given an entangled state, is there any explicit relation between measures of entanglement and the Riemannian metrics}? In this article, we focus exactly on this question, because we think that answering this question will enable us to gain some insights which will further help to find out unique Riemannian metrics corresponding to different measures of entanglement. To find out the answer of the question we consider two measures of entanglement, namely, concurrence($\mathcal{C}$)\cite{whooter} and negativity($\mathcal{N}$) \cite{Sanpera,Vidal}. We take different class of bi-partite entangled states whose sub-systems are non-maximally mixed qubits. Then we determine a particular Riemannian metric corresponding to those states using a theorem proposed by  Morozova and \u{C}encov.  Finally, we show that negativity and concurrence of such states are explicitly related to their corresponding Riemannian metrics. 

\paragraph*{} The rest of the article is arranged as follows. In Section(\ref{sec2})  we first provide an overview on Riemannian metric and Riemannian metrics on matrix space(quantum state space). Then we discuss an important theorem  on Riemannian metric. Section(\ref{sec3}) is dedicated to show our results.
In Section(\ref{sec4}) we conclude our work with discussions.

\section{Riemannian geometry of quantum state space}\label{sec2}
Riemannian geometry is a branch of differential geometry  which includes Riemannian manifolds and Riemannian metrics. Riemannian manifold is a real and smooth differentiable manifold embedded with an inner product at each point of the tangent space  and the inner product varies smoothly from point to point. More precisely, if $M$ is a differentiable manifold, $X$ and $Y$ are two vectors on the tangent space $T_xM$ passing through $x$ and $g_x$ is the inner product on the tangent space  at each point $x$,  then $x\mapsto g_x\{X(x),Y(x)\}$ is a smooth function. Riemannian metric on a manifold $M$ is the family of $g_x$. 
Morozova and \u{C}encov initiated the study of of monotone Riemannian metrics on the space of matrices. The motivation behind their work was to extend the geometric approach to quantum setting. They proposed the problem of finding Riemannian monotone metrics on the quantum state space which is endowed with a metric structure. 
\paragraph*{}
Quantum state space is identified with the set $\mathcal{M}_n$ of positive $n\times n$ matrices of trace one; they are termed as density matrices. This space of density matrices forms differential manifold on which a differentiable metric determines a Riemannian metric. On the other hand, the operators that act on the quantum states are expressed by $n\times n$  complex Hermitian matrices. The space of quantum operators is an inner product space, and the simplest inner product is the Hilbert-Schmidt one, defined as 
\begin{equation}
\label{1}
 \langle X,Y \rangle = \mbox{Tr} (X^*Y)
\end{equation}
where Tr is as usual matrix trace and $X,Y\in \mathcal{B}_n(\mathbb{C})$, $\mathcal{B}_n(\mathbb{C})$ being the set of complex self-adjoint matrices. This inner product is unitarily invariant that is 
$\langle X,Y \rangle = \langle UXU^\dagger, UYU^\dagger \rangle$ for every unitary $U$.
This invariance property is so strong that it determines the Hilbert-Schmidt inner product up to a constant multiple. 
\paragraph*{}
Now, by making the inner-products depending on quantum states($\rho$),  Riemannian monotone metric can be determined on the quantum state space
in the following way. Assume that for every $A,B\in \mathcal{B}_n(\mathbb{C})$, for every $\rho\in\mathcal{M}_n$, and for every $n\in N $, a complex quantity
$K_{\rho}(A,B)$ is given. The complex
quantity $K_{\rho}(A,B)$ will be a metric if the following conditions hold \cite{Pet1}:
\begin{enumerate}
 \item[(a)] $(A,B)\mapsto K_{\rho}(A,B)$ is sesquilinear.
 \item[(b)] $ K_{\rho}(A,A)\geq 0$, and the equality holds \emph{iff} $A=0$
 \item[(c)] $\rho\mapsto K_{\rho}(A,A)$ is continuous on $\mathcal{M}_n$ for every $A$
\end{enumerate}

The family of $K_{\rho}(A,B)$ with the above mentioned properties constitute  Riemannian metric on the differentiable manifold formed by the density matrices.
The Riemannian metric will be monotone if 
\begin{enumerate}
 \item [(d)]Under completely positive trace preserving(CPTP) map
$K_{\rho}(A,A)$ is contractive, i.e. $K_{\Lambda(\rho)}(A,A)\leq K_{\rho}(A,A)$ for every $\Lambda$, $\rho$ and $A$; 
$\Lambda(\cdot)$ being the CPTP map.
\end{enumerate}
For clear illustration of the metric $K_{\rho}(A,B)$, it is important to focus on the geometry of the quantum state space. Here $\mathcal{M}_n$ is the differential manifold and the self-adjoint operators $A$ and $B$ are the tangent vectors on the tangent space $T_{\rho}$. Therefore, $K_{\rho}(A,B)$ is the inner product on the tangent space $T_{\rho}$ at point $\rho$ . Considering the quantum state space to be finite dimensional, let us denote the set of all Hertmitian operators by  
\begin{equation}
\label{2}
 \mathcal{A}=\{A\lvert A=A^*\}
\end{equation}
and from the definition of $\mathcal{M}_n$;
\begin{equation}
\label{3}
 \mathcal{M}_n= \{\rho \lvert \rho=\rho^* \geq 0~ \mbox{and} ~\mbox{Tr} \rho=1\}
\end{equation}
The tangent space $T_{\rho}(\mathcal{M}_n)$ of each point $\rho$ may then be identified with
\begin{equation}
\label{4}
 \mathcal{A}_0= \{A\lvert A \in \mathcal{A} ~ \mbox{and}~ \mbox{Tr}A=0\}
\end{equation}
It can be shown that if $\mathcal{K}$ is an operator and $\mathcal{K}\in \mathcal{A}$, then $i [\rho, \mathcal{K}]$ will be an ordinary element of the tangent space $T_{\rho}(\mathcal{M}_n)$, that is, $i [\rho, \mathcal{K}]\in\mathcal{A}_0$\cite{Amari}. Therefore, by identifying tangent vectors Riemannian metric can be defined on the differential manifold formed by the density matrices, and if the metric satisfies condition (d) then the metric will be called monotone Riemannian metric.
\paragraph*{}
 Morozova and \u{C}encov tried to describe monotone metrics on the space of self-adjoint matrices but they were unable to show any metric. However, they proposed several candidates and provided an useful theorem. Later, Petz and other authors were able to find monotone metrics by introducing operator montone functions. Their works showed that there is an abundance of montone metrics on the space of self-adjoint 
matrices~\cite{Pet2,Pet3}. For our purpose, we will make use of the theorem provided by Morozova and \u{C}encov
\\\\\
THEOREM~\cite{Cencov,Pet1}:\emph{Assume that for every $D\in\mathcal{M}_n$ a real bilinear form $K_D^\prime$ is given on
the n-by-n self-adjoint matrices such that the conditions} (b),(c) and (d) \emph{are satisfied for self-adjoint A. 
Then there exists a positive continuous
function c($\lambda$,$\mu$) and a constant C with the following property: If D is diagonal with respect to the matrix units
$E_{ij}$, i.e. $ D = \sum_i \lambda_i E_{ii}$, then}

\begin{equation}
\label{5}
K^\prime(A,A)= C\sum_{i=1}^n\lambda_i^{-1}A_{ii}^2 + 2\sum_{i<j} \lvert A_{ij} \lvert ^2 c(\lambda_i, \lambda_j).
\end{equation}
\emph{for every self-adjoint $A=(A_{ij})$. Moreover if c is symmetric in its two variables, $c(\lambda,\lambda)= C \lambda^{-1}$
 and $c(t\lambda,t\mu)= t^{-1}c(\lambda,\mu)$ }.
 \\\\\
 
The function $c(\lambda,\mu)$ is termed as \emph{Morozova-\u{C}encov} function. The theorem tells that when $\mathcal{M}_n$ is considered as a differentiable manifold, the Riemannian metric must be a real bilinear form and the tangent vectors may be identified with self-adjoint matrices. Moreover, for all $D$ and for all self-adjoint operator $A$ the metric $K^\prime(A,A)$ can be determined using the above theorem.

\section{Results}\label{sec3}
It is already mentioned that in\cite{pd} an explicit relation between negativity and Riemannian metric has been established. In brief, if a qubit $\rho_S$ interacts with an ancillary qubit $\rho_A$ and an
entangled state $\rho_{SA}$ is produced, then negativity $\mathcal{N}$ of the entangled state is, upto a constant factor, equal to the square root of the Riemannian metric defined for $\rho_S$. A specific unitary $U_{SA}$ produce the entangled state by acting on the initial product state $\rho_S\otimes\lvert 0\rangle_A$. Here we will show that for different class of entangled states explicit relations between some measures of entanglement and Riemannian metric exist. To begin with, let us consider an entangled state $\rho_{A_1A_2}$ consisting of subsystems $A_1$ and $A_2$. Negativity \cite{Vidal} of this state is given by 
\begin{equation}
\label{6}
 \mathcal N(\rho_{A_1A_2})=\frac{{\lvert\lvert\rho_{A_1A_2}^{T_{A_1}}\rvert\rvert}_1-1}{2}
\end{equation}
where $\rho_{A_1A_2}^{T_{A_1}}$ denotes the partial transpose with respect to the subsystem $A_1$ and ${\lvert\lvert\rho_{A_1A_2}^{T_{A_1}}\rvert\rvert}_1$ denotes the trace norm of the matrix. 
Concurrence \cite{whooter} of the state $\rho_{A_1A_2}$ is given by
 \begin{equation}
 \label{7}
  \mathcal{C}=\max\{0,\lambda_1-\lambda_2-\lambda_3-\lambda_4\}
 \end{equation}
where $\lambda's$ are the square root of of the eigenvalues of $\rho_{A_1A_2}\tilde\rho_{A_1A_2}$ ~in decreasing order. The spin-flipped density matrix 
$\tilde\rho_{A_1A_2}$ is defined as
\begin{equation}
\label{8}
 \tilde\rho_{A_1A_2}=\sigma_y^{A_1}\otimes\sigma_y^{A_2}\rho^*\sigma_y^{A_1}\otimes\sigma_y^{A_2}
\end{equation}
where * denotes the complex conjugate in the computational basis. In order to establish explicit relation between these two measures and Riemannian metric, we need to determine Riemannian metric corresponding to any of the subsystem state. For this purpose we will use the theorem mentioned earlier. Now we consider different class of entangled states and show the proposed results.
 
\subsection{Pure entangled state}
 Let us take a pure two qubit entangled state 
\begin{equation}
\label{9}
 |\psi\rangle_{A_1A_2}=\alpha|00\rangle+\beta|11\rangle
\end{equation}
where $ \lvert\alpha\rvert^2+\lvert\beta\rvert^2=1$. The density matric corresponding to this state is $\rho_{A_1A_2}$. Tracing out subsystem $A_2$ we get the state $\rho_{A_1}$ of the subsystem $A_1$ as:
\begin{equation}
\label{10}
\begin{split}
 \rho_{A_1} & =\mbox{Tr}_{A_2}(\rho_{{A_1}{A_2}})\\
 & = \lvert\alpha\rvert^2|0\rangle\langle0|+\lvert\beta\rvert^2|1\rangle\langle1|
\end{split}           
\end{equation}
Our aim is to find the metric
$K_{\rho_{A_1}}(A,B)$. From the definition of the metric $A$ and $B$ are traceless self-adjoint operators and they act as tangent vectors corresponding to the Riemannian manifold formed by single qubit density matrices. In order to construct $A$ and $B$ with their respective properties we use the fact that the Pauli matrices are self-adjoint traceless operators and define them as $A=B=\mbox{i}[\rho_{A_1},\sigma_l\mid l=x,y,z]$, where `$\mbox{i}$' stands for imaginary.
 Matrix representation of these operators are given by:
\begin{equation}
\label{11}
\mbox{i}[\rho_{A_1},\sigma_x]=\mbox{i}\begin{pmatrix} 0 &(\lvert\alpha\rvert^2-\lvert\beta\rvert^2) \\ (\lvert\beta\rvert^2-\lvert\alpha\rvert^2)& 0 \end{pmatrix}
\end{equation}
and
\begin{equation}
\label{12}
\mbox{i}[\rho_{A_1},\sigma_y]=\begin{pmatrix} 0 &(\lvert\alpha\rvert^2-\lvert\beta\rvert^2) \\ (\lvert\alpha\rvert^2-\lvert\beta\rvert^2)& 0 \end{pmatrix}
\end{equation}
It is easy to verify from Eq.(\ref{11}) and Eq.(\ref{12}) that trace of the two self-adjoint operators are zero. As the diagonal elements
of the matrix corresponding to the operators $\mbox{i}[\rho_{A_1},\sigma_x]$ and $\mbox{i}[\rho_{A_1},\sigma_y]$ are zero i.e. $A_{11}$=$A_{22}$=0,
 the first summation term of the Eq.(\ref{5}) is zero i.e
 \begin{equation}
 \label{13}
  C\sum_{i=1}^2\lambda_i^{-1}A_{ii}^2=0
 \end{equation}
 Therefore, for operator $i[\rho_{A_1},\sigma_x]$ we get  Riemannian metric as:
 \begin{equation}
 \label{14}
  K_{\rho_{A_1}}(\mbox{i}[\rho_{A_1},\sigma_x],\mbox{i}[\rho_{A_1},\sigma_x])=2\sum_{i<j}\lvert A_{ij}\rvert^2c(\lambda_i,\lambda_j)
 \end{equation}
 where $A_{ij}$ are the off-diagonal elements of the matrix given in Eq.(\ref{11}) and $c(\lambda_i,\lambda_j)$ is the Morozova-\u{C}encov function
 \cite{Cencov,Pet1}. For our calculation, we take one of the functions proposed originally by Morozova and \u{C}encov\cite{Cencov}:
\begin{equation}
\label{15}
 c(\lambda_i,\lambda_j)=\frac{2}{\lambda_i+\lambda_j}.
\end{equation}
 Using Eq.(\ref{11}) and Eqs.(\ref{13},\ref{14},\ref{15}) we get the Riemannian metric $K_{\rho_{A_1}}(\mbox{i}[\rho_{A_1}\sigma_x],\mbox{i}[\rho_{A_1},\sigma_x])$ corresponding to the state $\rho_{A_1}$ as
\begin{equation}
\label{18}
\begin{split}
 K_{\rho_{A_1}}(\mbox{i}[\rho_{A_1},\sigma_x],\mbox{i}[\rho_{A_1},\sigma_x]) &= 2 |A_{12}|^2 \frac{2}{\lambda_1+\lambda_2}\\
 & =2(|\alpha|^2-|\beta|^2)^2\frac{2}{|\alpha|^2+|\beta|^2}\\
 & = 4(1-4|\alpha|^2|\beta|^2)
\end{split}
\end{equation}
where $\lambda_1$ and $\lambda_2$ are eigenvalues of the density matrix $\rho_{A_1}$ and $A_{12}$ is the off-diagonal element of the 
operator $\mbox{i}[\rho_{A_1},\sigma_x]$.  Now using Eqs.(\ref{6},\ref{7}) we get the negativity and concurrence of the state $\rho_{{A_1}{A_2}}$ as
 \begin{equation}
 \label{16}
  \mathcal N(\rho_{A_1A_2})=|\alpha||\beta|
 \end{equation}
 and
 \begin{equation}
 \label{17}
  \mathcal{C}=2|\alpha||\beta|.
 \end{equation}
\\
Finally, from Eq.(\ref{18}) and Eqs.[\ref{16},\ref{17}] we get 
\begin{equation}
\label{19}
\begin{split}
 K_{\rho_{A_1}}(\mbox{i}[\rho_{A_1},\sigma_x],\mbox{i}[\rho_{A_1},\sigma_x]) & = 4(1-4\mathcal{N}^2)\\
 & = 4(1-\mathcal{C}^2)
\end{split}
\end{equation}
The above equation is valid for $\sigma_y$ as well, i.e., 

\begin{equation}
\label{20}
\begin{split}
 K_{\rho_{A_1}}(\mbox{i}[\rho_{A_1},\sigma_y],\mbox{i}[\rho_{A_1},\sigma_y]) & = 4(1-4\mathcal{N}^2)\\
 & = 4(1-\mathcal{C}^2)
\end{split}
\end{equation}
These two equations represent explicit relation between measures of quantum correlations(entanglement) and Riemannian metric for  two qubit pure entangled state. Due to unitary invariance property of concurrence, negativity and Riemannian metric, the above equations are also unitary invariant. In the next subsection we will consider mixed entangled states.  

\subsection{Mixed entangled state}\label{}
In this section we are going to show the relation between the Riemannian metric and measures of entanglement for two-qubit mixed entangled state. We consider two qubit maximally entangled mixed 
state(MEMS)\cite{ishizaka} and non-maximally entangled mixed state. 

\subsubsection{Maximally Entangled Mixed State}
Maximally entangled mixed states(MEMS) are those states that have the maximum possible entanglement for a given mixedness. MEMS were first introduced by Ishizaka and Hiroshima\cite{ishizaka}in a way that 
their entanglement is maximized by fixing the eigenvalues of the density matrices. The amount of entanglement of these states can not be increased by any global unitary transformation and this property will hold for states having rank less than 4. Later Munro et al. \cite{munro} had derived an analytical form of MEMS and showed that these states are optimal for the entanglement(concurrence) and purity measure. Wei et.al \cite{wei} further showed that MEMS depend on the measures one uses to quantify entanglement. For different entanglement measures and mixedness there are different form of maximally
entangled mixed state (MEMS). Here we will show relation between Riemannian metric and entanglement measures for different form of maximally
entangled mixed state (MEMS).

\paragraph{MEMS of Ishizaka and Hiroshima:}

Ishizaka and Hiroshima have proposed maximally entangled mixed states whose entanglement is maximized at fixed eigenvalues 
for different rank.
\begin{itemize}
 \item Rank-4 state:\newline
 The states they have proposed are those which can be obtained by applying any local unitary transformation on the state
 \begin{eqnarray}
 \label{22}
  \rho_{A_1A_2}&=&p_1|\psi^-\rangle\langle\psi^-|+p_2|00\rangle\langle00|\nonumber\\
  &&+p_3|\psi^+\rangle\langle\psi^+|+p_4|11\rangle\langle11|
 \end{eqnarray}
where $|\psi^{\pm}\rangle=(|01\rangle\pm|10\rangle)/\sqrt{2}$ are the Bell states and $|00\rangle$, $|11\rangle$ are product states 
orthogonal to $|\psi^{\pm}\rangle$. $p_i$ are the eigenvalues of $\rho_{A_1A_2}$ in decreasing order $(p_1\geq p_2\geq p_3
\geq p_4)$ and $p_1+p_2+p_3+p_4=1$. Ishizaka and Hiroshima have shown that Rank-4 states will be MEMS if the following relation is satisfied:
\begin{equation}
\label{29}
p_3=p_2+p_4-\sqrt{p_2p_4}
\end{equation}
Now, concurrence of the state given in Eq.(\ref{22}) is found to be
\begin{equation}
\label{30}
\mathcal{C}=p_1-p_3-2\sqrt{p_2p_4}
\end{equation}

The state of subsystem $A_1$ is
\begin{eqnarray}
\label{23}
 \rho_{A_1}&=&\big(\frac{p_1+p_3}{2}+p_2\big)|0\rangle\langle0|\nonumber\\
 &&+\big(\frac{p_1+p_3}{2}+p_4\big)|1\rangle\langle1|
\end{eqnarray}
 Riemannian metric corresponding to this state will be 
\begin{eqnarray}
\label{27}
K_{\rho_{A_1}}(\mbox{i}[\rho_{A_1},\sigma_x],\mbox{i}[\rho_{A_1},\sigma_x]) & =&2\sum_{i<j}\lvert A_{ij}\rvert^2c(\lambda_i,\lambda_j)\nonumber\\
& =&4(p_2-p_4)^2
\end{eqnarray}
where $c(\lambda_i,\lambda_j)=2/(\lambda_i+\lambda_j)$.

\/*
Now we introduce a quantity called $linear~ entropy$ which is a measure of mixedness of any state. For N-dimensional state linear entropy can be written as
\begin{equation}
 \label{31}
 S_L=\frac{N}{N-1}(1-\mbox{Tr}(\rho^2))
\end{equation}
which ranges from 0 (for pure state) to 1 (maximally mixed state).
Using the above formula we get the linear entropy of the rank-4 state as
\begin{eqnarray}
 \label{32}
 S_L&=&\frac{4}{3}(1-p_2^2-p_4^2)-\frac{(p_3-p_1)^2}{2}\nonumber\\
 &&-\frac{(p_1+p_3)^2}{2})
\end{eqnarray}
*/
 Using Eqs.[\ref{29},\ref{30},\ref{27}] we finally get the relation between concurrence and Riemannian metric as
\begin{equation}
 \label{33}
 \sqrt{K_{\rho_{A_1}}(\mbox{i}[\rho_{A_1},\sigma_x],\mbox{i}[\rho_{A_1},\sigma_x])}
=\frac{2}{3}(1-\mathcal{C})-4p_4
 \end{equation}

\item Rank-3 state: \newline
MEMS of rank-3 can be derived from Eq.(\ref{22}) by putting $p_4=0$:
\begin{eqnarray}
\label{35}
\rho_{A_1A_2}&=&p_1|\psi^-\rangle\langle\psi^-|+p_2|00\rangle\langle00|\nonumber\\
&&+p_3|\psi^+\rangle\langle\psi^+|
\end{eqnarray}
where $p_1+p_2+p_3=1$. After doing similar calculations as in previous case we obtain the relation between Riemannian metric and concurrence as
\begin{equation}
 \label{36}
\sqrt{K_{\rho_{A_1}}(\mbox{i}[\rho_{A_1},\sigma_x],\mbox{i}[\rho_{A_1},\sigma_x])}
=2(1-\mathcal{C})-4p_3
\end{equation}

\item Rank-2 state: \newline
Putting $p_3=p_4=0$ in Eq.(\ref{22})  we get rank-2 MEMS
\begin{equation}
 \label{39}
 \rho_{A_1A_2}=p_1|\psi^-\rangle\langle\psi^-|+p_2|00\rangle\langle00|
\end{equation}
where $p_1+p_2=1$. The relations between concurrence and Riemannian metric for such states is found to be
\begin{equation}
 \label{40}
\sqrt{K_{\rho_{A_1}}(\mbox{i}[\rho_{A_1},\sigma_x],\mbox{i}[\rho_{A_1},\sigma_x])}
 =2(1-\mathcal{C})
\end{equation}

\paragraph*{} In deriving the relations between Riemannian metric and concurrence for states of different rank we have considered subsystem $A_1$. However, similar relations can also be derived if we take 
subsystem $A_2$ as the density matrices of two subsystems are same.

\end{itemize}

 \paragraph{MJW MEMS:}
 
 Wei et al.\cite{wei} had shown that MEMS have different form for different entanglement measures. They had derived analytical form of MEMS for
 different entanglement measures for a given amount of mixedness. Here we have taken a MEMS for negativity measure 
 \begin{equation}\label{45}
 \begin{split}
 \rho_{A_1A_2}&=\left(\frac{1+\sqrt{3r^2+1}}{6}\right)|00\rangle\langle00|+\frac{r}{2}|00\rangle\langle11|\\
 &+\frac{r}{2}|11\rangle\langle00|+\left(\frac{1+\sqrt{3r^2+1}}{6}\right)|11\rangle\langle11| \\
 &+\left(\frac{4-2\sqrt{3r^2+1}}{6}\right)|01\rangle\langle01|
 \end{split}
  \end{equation}
where $0\leq r\leq1$. Now the state of subsystem $A_1$ is
\begin{equation}\label{46}
\rho_{A_1}=\left(\frac{5-\sqrt{3r^2+1}}{6}\right)|0\rangle\langle0|+\left(\frac{1+\sqrt{3r^2+1}}{6}\right)|1\rangle\langle1|
 \end{equation}
After calculating negativity and Riemannian metric, we finally get the relation between them as 
\begin{equation}\label{47}
 \sqrt{K_{\rho_{A_1}}(\mbox{i}[\rho_{A_1},\sigma_x],\mbox{i}[\rho_{A_1},\sigma_x])}
 = \frac{4}{3}(1-\mathcal{N})
\end{equation}

\subsubsection{Non-maximally Entangled Mixed State}
 We have considered maximally entangled mixed state so far. Now let us take non-maximally entangled mixed state. One such example is the state 
\begin{equation}\label{43}
 \rho_{A_1A_2}=p|\psi\rangle\langle\psi|+(1-p)|01\rangle\langle01|
\end{equation}
 where $|\psi\rangle=\alpha|00\rangle+\beta|11\rangle$ is a non-maximally entangled pure state and
$|\alpha|^2+|\beta|^2=1$.
For this state we get the relation between Riemannian metric and concurrence as
\begin{equation}\label{44}
 \sqrt{K_{\rho_{A_1}}(\mbox{i}[\rho_{A_1},\sigma_x],\mbox{i}[\rho_{A_1},\sigma_x])}=
 2(1-2\frac{\alpha}{\beta}\mathcal{C})
\end{equation}

\section{Conclusion} \label{sec4}
Quantum correlation is a resource for quantum information processing tasks, and entanglement is the best studied form of it. However, till date, no attempt has been taken to study quantum entanglement using geometry of quantum state space. In this article we have addressed a problem, which we think, can shed some light on the study of quantum entanglement using Riemannian metrics on quantum state space. We have considered different class of two qubit entangled states and for each class we have shown an explicit relation between measures of entanglement and Riemannian metric. The measures that we have taken into consideration are negativity and concurrence. Riemannian metric on the differential manifold has been constructed by using a theorem provided by Morozova and \u{C}encov. The entangled states that we have considered have a common feature; the subsystems are non-maximally mixed. For maximally mixed subsystems the Riemannian metric will be zero. Therefore, in such cases we can not get explicit relation between measures of entanglement and Riemannian metric. This is infact a limitation of the process. Though the results of this paper do not mention any unique mapping between Riemannian metric and measures of entanglement, they certainly emphasize that there exist explicit relation between measures of quantum correlation and geometry of quantum state space. We hope that our work will be useful in defining unique Riemannian metrics corresponding to different entanglement measures.\\
$\bf {Acknowledgement:}$
The authors would like to acknowledge Scientific and Engineering Research Board, Govt. of India for financial support.


\end{document}